\documentclass[%
 rsi,
 amsmath,amssymb,
 preprint,
]{revtex4-1}

\usepackage{graphicx}
\usepackage{bm}
\usepackage[utf8]{inputenc}
\usepackage[T1]{fontenc}
\usepackage{mathptmx}
\usepackage{etoolbox}
\usepackage{units}
\usepackage{color}
\makeatletter

\def\@email#1#2{%
 \endgroup
 \patchcmd{\titleblock@produce}
  {\frontmatter@RRAPformat}
  {\frontmatter@RRAPformat{\produce@RRAP{*#1\href{mailto:#2}{#2}}}\frontmatter@RRAPformat}
  {}{}
}%
\makeatother

\begin{document}
\title[]{A spatially resolved optical method to measure thermal diffusivity}
\author{F. Sun}
\author{S. Mishra}
\author{P. H. McGuinness}
\author{Z. H. Filipiak}
\author{I. Marković}
\author{D. A. Sokolov}
\affiliation{ Max Planck Institute for Chemical Physics of Solids, 01187 Dresden, Germany}

\author{N. Kikugawa}
\affiliation{National Institute for Materials Science, Ibaraki 305-0003, Japan}

\author{J. W. Orenstein}
\affiliation{Department of Physics, University of California, Berkeley, California 94720, USA}
\affiliation{Materials Science Division, Lawrence Berkeley National Laboratory, Berkeley, California 94720, USA }

\author{S. A. Hartnoll}
\affiliation{Department of Applied Mathematics and Theoretical Physics, University of Cambridge, Cambridge CB3 0WA, UK}

\author{A. P. Mackenzie}
\affiliation{ Max Planck Institute for Chemical Physics of Solids, 01187 Dresden, Germany}
\affiliation{School of Physics and Astronomy, University of St Andrews, St Andrews KY16 9SS, UK}

\author{V. Sunko*}
\affiliation{ Max Planck Institute for Chemical Physics of Solids, 01187 Dresden, Germany}
\affiliation{Department of Physics, University of California, Berkeley, California 94720, USA}

\email{The author to whom correspondence may be addressed: vsunko@berkeley.edu}
\date{\today}

\begin{abstract}

We describe an optical method to directly measure position-dependent thermal diffusivity of reflective single crystal samples across a broad range of temperatures for condensed matter physics research. Two laser beams are used, one as a source to locally modulate the sample temperature, and the other as a probe of sample reflectivity, which is a function of the modulated temperature. Thermal diffusivity is obtained from the phase delay between source and probe signals. We combine this technique with a microscope setup in an optical cryostat, in which the sample is placed on a 3-axis piezo-stage, allowing for spatially resolved measurements. Furthermore, we demonstrate experimentally and mathematically that isotropic in-plane diffusivity can be obtained when overlapping the two laser beams instead of separating them in the traditional way, which further enhances the spatial resolution to a micron scale, especially valuable when studying inhomogeneous or multidomain samples. We discuss in detail the experimental conditions under which this technique is valuable, and demonstrate its performance on two stoichiometric bilayer ruthenates: Sr$_{3}$Ru$_{2}$O$_{7}$ and Ca$_{3}$Ru$_{2}$O$_{7}$. The spatial resolution allowed us to study the diffusivity in single domains of the latter, and we uncovered a temperature-dependent  in-plane diffusivity anisotropy. Finally, we used the enhanced spatial resolution enabled by overlapping the two beams to measure temperature-dependent diffusivity of Ti-doped Ca$_{3}$Ru$_{2}$O$_{7}$, which exhibits a metal-insulator transition. We observed large variations of transition temperature over the same sample, originating from doping inhomogeneity, and pointing to the power of spatially resolved techniques in accessing inherent properties. 
 
\end{abstract}

\maketitle

\section{\label{sec:Intro}Introduction}
Thermal transport experiments are one of the key tools used in condensed matter physics to understand fundamental physics of electrons, phonons and their interactions in correlated materials over a wide range of temperatures. Thermal conductivity ($\kappa$) is considered as a characteristic quantity, and is widely studied to explore the transport behavior and scattering mechanisms \cite{allen1994thermal,minami2003influence,checkelsky2012thermal}. Combined with measurements of the electrical resistivity, it allows construction of the Wiedemann-Franz ratio \cite{kumar1993experimental}, which provides important information on electron and phonon scattering. However, thermal radiation typically limits the measurement of $\kappa$ performed in cryostats to the low-temperature regime (below 100 K in most cases); measurements at higher temperatures require specifically dedicated setups in which special care must be taken to calibrate and correct for radiation losses, induced by the difference in temperature between the sample and the surrounding cryostat, which is usually kept at the temperature of liquid helium (\unit[4]{K}) \cite{talpe1991reduction}. One of the consequences of this experimental limitation is that there are only a few measurements of thermal transport in the wide temperature range over which the universal `Planckian’ scattering rate has been observed in electrical transport measurements \cite{zaanen2004temperature,bruin2013similarity,legros2019universal,mousatov2020planckian,mousatov2021phonons}. 

Thermal conductivity is usually determined by measuring a temperature difference induced by a known heat flux across a sample thermally insulated from the environment. Such an approach has many advantages, notably its easy implementation in low-temperature setups and the possibility to measure electrical transport using the same contacts. Therefore, geometric uncertainty is greatly reduced when evaluating the ratio of thermal and electrical conductivity. However, this method requires  samples to be homogeneous over at least several hundreds of microns in length, and thermally isolated from the environment. Furthermore, investigations of thermal transport anisotropy require preparation of several samples with different contact geometries, a laborious process introducing the possibility of systematic errors. While the above limitations can be overcome in large homogeneous single crystals, they are likely to introduce errors in materials that are inhomogeneous, or in symmetry-broken phases exhibiting spontaneous domain formation. In addition, the same limitations become prohibitive in thin films which cannot be thermally insulated from substrates they were grown on, and in van der Waals hetero-structures, which are typically only several tens of microns in size. 

The above considerations motivate a different approach to study the thermal transport.  We were particularly influenced by recent work \cite{hartnoll2015theory,behnia2019lower,Zhang5378,zhang2020optical}, suggesting that thermal diffusivity ($D$), instead of thermal conductivity ($\kappa$), can be a more informative quantity to describe thermal transport in the high-temperature regime, at least in transient conditions. Traditionally, thermal diffusivity is obtained from separate measurements of $\kappa$ and heat capacity ($c$) ($D$ = $\kappa$ / $c$), therefore the limitations of the thermal conductivity measurements are inherited, as well as the acquired uncertainties related to the two separate measurements. Real samples also frequently contain inhomogeneities, either chemical or due to the formation of domains in ordered states of interest, so a technique with high spatial resolution is desirable.  Extensive work in several fields of research has shown that this can be achieved optically, using two laser beams, one to locally heat the sample and a second to probe the response to that heating.  If the heat is applied at non-zero frequency, the relative phase of the probe and heat signals contains information on the thermal diffusivity.  Techniques based on this principle, known as modulated thermal reflectance microscopy or photothermal reflectance microscopy, are extensively used in materials science, and have been refined to produce compact bench-top instruments for the room-temperature characterisation of materials \cite{bocchini2016thermal,potenza2017graphene,fabbri1996analysis,fournier2020measurement,matsui2011analysis,hartmann1997measuring}.  For the physical problems in condensed matter physics outlined above, however, room-temperature measurements are not sufficient.  It is necessary to combine the optics with cryogenics, producing an instrument enabling measurement over a large temperature range.  Further, it is desirable to incorporate micron-scale spatial resolution, for example to enable work with inherently inhomogeneous samples.  In this paper we describe our approach to the design of such an instrument and its use.

Direct measurement of $D$ using lasers started from a flash method at high temperatures in 1960s, which records the temperature versus time over a centimeter-scale sample after heating it up with a light pulse  \cite{parker1961flash,rudkin1962thermal}. This method relied on a direct measurement of the value of temperature increase due to the laser heating, making it difficult to account for thermal losses and impedance between interfaces, limitations similar to those experienced using contact methods. Furthermore, measurement of semi-transparent
materials can lead to other problems, solvable with special tricks \cite{bellucci2019transmittance, potenza2021numerical}. From the 1990s, a photo-deflection method has been applied to measure $D$ without direct temperature measurement, by monitoring changes to the refractive index of a gas above the sample surface, but the technique still requires a mm scale sample, and is unsuitable for cryogenic applications extending below the boiling point of nitrogen at 77 K \cite{salazar1991thermal,salazar1993thermal,bertolotti1993photodeflection}. The required sample size was further reduced to $\sim$50 µm by employing two laser spots, one as a source of heat and one as a probe of sample reflectivity \cite{wu1993photothermal}. This principle was also adopted in materials science, leading to the development of the room-temperature microscopies mentioned above. For condensed matter physics applications, cryostats can be incorporated, enabling its application to quantum materials \cite{wu1993photothermal,Zhang5378,ThermalPRB2019,zhang2020optical}.

To maximise the suitability of the laser-based technique to quantum materials research, it is desirable to simultaneously maximise spatial resolution and spatial range (the area over which scanning can be performed) over a wide range of temperatures.  In order to achieve this, we build on the previous work discussed above, introducing a number of modifications based on technologies recently developed for cryogenic optics experiments.  We operate in vacuum in a dedicated optical cryostat (Montana Instruments Cryostation S50) and obtain high range and resolution by using a piezo-driven scanning stage to move the sample.  Although we can also move the laser spots, this occurs over a much more limited range, and most experiments are done in a moving-sample-fixed-optics mode.  To profit from the high spatial resolution offered by the scanning stage, we mathematically and experimentally demonstrate that the in-plane diffusivity can be measured by overlapping the source and probe beams. Combining these approaches yields an instrument capable of scanning regions $\sim$ 4000 $\times$ 4000 µm, with $\sim$ 1 µm spatial resolution at temperatures ranging from 5 K to 330 K. Furthermore, we describe in detail how to minimise artefacts arising from thermal contraction of the setup, and detail the parameter range over which reliable quantitative measurements of diffusivity are possible.

\section{\label{sec:ExpandTeo} Experimental design}

\subsection{\label{sec:Exp} The setup}

\begin{figure*}
\includegraphics[width=6in]{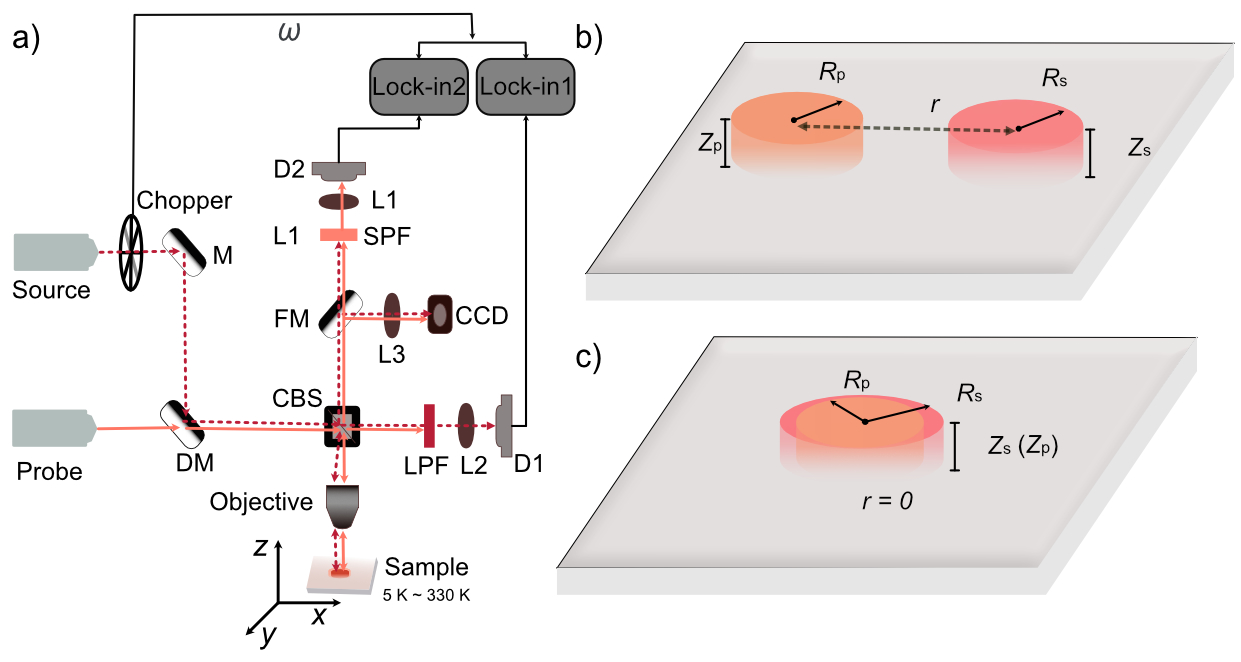}
\caption{\label{Fig1} The schematic of the optical experiments. a) the setup of the optical path. M: mirror; L: lens; LPF: long pass filter; SPF: short pass filter; CBS: cubic beam splitter; FM: flip mirror; D: photo diode; DM: dichroic mirror. The sample is placed inside a cryostat, which allows for temperature-dependent measurements in the \unit[5-330]{K} range. The $x$, $y$, and $z$ represent the three axes of the piezo-stage. b) Separated configuration. c) Overlapped configuration.}
\end{figure*}

Our optical setup for thermal diffusivity measurements is shown in FIG. \ref{Fig1}(a). Two laser beams are used, one as a source ($\unit[780]{nm}$) to locally increase the sample temperature, and the other as a probe ($\unit[633]{nm}$) of reflectivity, which is a function of sample temperature. Both source and probe beams are focused by the same objective onto the surface of the sample as spots of $\sim$2 µm radius. The source beam is modulated by a mechanical chopper, causing periodic heating at the frequency selected by the chopper. Sample temperature, detected by the probe beam, changes at the same frequency. However, the temperature change is not instantaneous; the time scale at which it occurs depends on thermal diffusivity, which can therefore be obtained from the phase delay between source and probe signals. In order to measure the phase difference between the source and probe, two lock-in amplifiers are used. One records the source and the other one measures the probe signal, both sharing the same modulation frequency of the optical chopper. Measurements are performed in two configurations, which we refer to as separated and overlapped: in the former (used in the previous work in the field that we reviewed in Section I above) the two laser spots are separated by a distance $r$ (FIG. \ref{Fig1}(b)), while in the latter they spatially overlap (FIG. \ref{Fig1}(c)). In FIGs. \ref{Fig1}(b, c) $R_s$ and $R_p$ represent the radius of the source and probe, $r$ is the distance between the center of the two spots (a finite value for separated case and 0 for overlapped case), and $z_s$ and $z_p$ represent the penetration depths of the source and probe lasers, respectively. 

The mathematical description of the experiment, valid in both configurations, is based on the anisotropic diffusion equation:
\begin{equation}
 \frac{\partial{\Phi}}{\partial t}-
 (D_x\frac{\partial^2{\Phi}}{\partial x^2}+D_y\frac{\partial^2{\Phi}}{\partial y^2}+D_z\frac{\partial^2{\Phi}}{\partial z^2}) = \frac{{I}}{c},
 \label{diffequ}
\end{equation}
where the $D_x$, $D_y$ and $D_z$ are the thermal diffusivities along three principal directions of the crystal, $c$ is the heat capacity, $\Phi(x,y,z,t)$ is the temperature profile, and $I(x,y,z,t)$ is the intensity of the external heat source. Since the diffusivity tensor is diagonal only in the basis defined by its principal axes, it is important to identify their orientation prior to performing measurements based on Eq.~\ref{diffequ}, as we do when determining the anisotropic diffusivity of orthorhombic Ca$_{3}$Ru$_{2}$O$_{7}$. We use two methods to identify the principal axes: spatial mapping of optical birefringence (FIG. \ref{Fig3}(b)), and a measurement of the angular dependence of diffusivity, where the maximum and minimum values correspond to the orientations of the principal axes (FIG. \ref{Fig3}(d)). The intensity profile of source ($I_s (x,y,z,t)$) and probe ($I_p (x,y,z,t)$) are given by the real part the following expressions: 
\begin{equation}
 I_s (x,y,z,t) \propto e^{-(x^2+y^2)/{2R_s^2}}e^{-z/z_s}e^{-i\omega t},
 \label{source}
\end{equation}
\begin{equation}
 I_p (x,y,z,t) \propto e^{-(x^2+y^2)/{2R_p^2}}e^{-z/z_p}e^{-i\omega t},
 \label{probe}
\end{equation}
where $\omega$ is the modulation frequency, $z_s$ and $z_p$ are the penetration depths of the light from the source and probe spots perpendicular to the sample surface, and $R_s$ and $R_p$ the spot radii of the two beams. By solving the diffusion equation (details can be found in Appendix A), we obtain the spatially and temporally resolved temperature profile, and therefore the phase delay between the excitation and probe. Here we discuss two experimental configurations.

\subsection{\label{sec:Sep}Separated configuration ($r \gg R_s {\rm ,} R_p$)}

In the separated case (FIG. \ref{Fig1}(b)), the typical values of the experimental parameters are $r$ $\approx$ 25 µm, $R_s$ $\approx$ $R_p$ $\approx$ 2 µm, and both the source and the probe spots (Eq.\ref{source}) can be approximated as delta functions. Since the spots are separated in the plane, it is clear that the experiment tests the in-plane diffusion, and the extent of the beams perpendicular to the plane can be neglected ($z_s = z_p= 0 $). The phase delay between source and probe is then given by:
\begin{equation}
 \phi = \sqrt{\frac{r^2\omega}{2D}},
 \label{SepPhase}
\end{equation}
where $D$ is the thermal diffusivity, $\omega$ is the modulation frequency, $r$ is the separation between source and probe spots, and $\phi$ is the phase difference \cite{Zhang5378,ThermalPRB2019}. The  phase delay $\phi$ is proportional to the square root of the modulation frequency ${\omega}$. Therefore, the validity of the model can be directly checked by measuring the frequency dependence of the phase.

\subsection{\label{sec:ol} Overlapped configuration ($r = 0$)}

The mathematical description of the novel overlapped configuration (FIG. 1(c)) is complicated by the fact that the Gaussian profile of the two beams needs to be explicitly taken into account. Furthermore, it is not immediately obvious whether this experiment probes the in-plane diffusion, the out-of-plane diffusion or a combination of the two. We numerically compare the solutions to Eq.~\ref{diffequ} taking into account the finite penetration depth with those obtained using a fully 2-dimensional approximation (Appendix B), and find that this experiment is dominated by the in-plane diffusion for realistic parameters. The underlying physical reason is the small size of penetration depth of the light compared to the typical thermal diffusion length; the temperature is consequently essentially constant across the probing depth, but it decreases along the radius of the beam. In the small frequency limit ($\omega \ll \frac{D}{R_s^2+R_p^2}$)  we find an analytical solution for the phase delay:

\begin{equation}
 \phi = \sqrt{\frac{(R_s^2+R_p^2)\omega}{\pi D}},
 \label{OlPhase}
\end{equation}
The phase delay $\phi$ is again proportional to $\sqrt{\omega}$, similarly to the separated configuration, but the characteristic length is now the spot size, rather than the larger spot separation. 

\subsection{\label{sec:config}Choice of configuration}
Each of the two measurement configurations (overlapped and separated) has advantages and disadvantages and may therefore be appropriate under different conditions. The larger relevant length scale in the separated case results in a larger phase delay, making it less sensitive to systematic errors such as laser fluctuations, small focal shifts, \textit {etc}. Furthermore, by controlling the direction of separation between the two spots with respect to crystalline axes it is possible to utilise the separated configuration to measure diffusivity anisotropy. In contrast, the overlapped case offers better spatial resolution, but is more sensitive to imperfections in beam overlap and changes of the focus. The latter is particularly problematic when measuring temperature dependence of diffusivity: due to the unavoidable thermal expansion of the sample holder, a 20-µm-shift in the vertical position of the sample is observed during temperature changes from 330 K to 50 K, which brings it out of focus along the $z$-axis. We were able to precisely calibrate and compensate that temperature-dependent shift (Appendix C) by adjusting the vertical position of the sample using piezo-stages. In addition, we designed two ways of switching between the two measurement configurations: one using a mirror and another requiring coordination between two mirrors. The former is easier to operate and is applicable to single-temperature experiments, while the latter is more stable against $z$-shifts of the sample stage during the temperature-dependent experiments. Details can be found in Appendix D. To summarise, the overlapped configuration offers better spatial resolution, but potentially suffers from more experimental artefacts, which must be accounted for by careful experimental design and calibration, which we describe in this paper. 

\subsection{\label{sec:SNR}Discussion of the fundamental experimental signal $\delta \rho $}

In both configurations, the inverse diffusivity is proportional to the square of the phase shift between the local temperature change $\delta T$ and the power source causing the temperature change. The fundamental signal measured in our experiments is the change of reflectivity caused by the temperature change, $\delta \rho $, which is proportional to $\delta T$ \emph{if} the heating is weak enough to keep the system in the linear response regime. In order to ensure that the measurements are performed under these conditions, it is important to measure $\delta \rho $ as a function of the source power (Appendix E), and choose a suitable power to remain in the linear regime. Special care needs to be taken in the vicinity of phase transitions, where both the reflectivity and the diffusivity might be changing rapidly with temperature; typically a lower power will need to be used compared with deep within a phase. Furthermore, different power levels may be appropriate at temperatures below and above a phase transition.   

This measurement scheme requires reflectivity to have a measurable temperature dependence. Since our modulation technique allows the detection of relative reflectivity changes in the order of $10^{-6}$, and the signal can be further increased by increasing the source power (while staying in the linear regime), we expect that, for large majority of materials, the thermal diffusivity can be measured via this optical method over extended temperature ranges. One exception is the unusual case in which the temperature derivative of reflectivity vanishes, \textit{e.g.} the reflectivity undergoes a maximum or minimum. Even in such a situation a measurement at a different wavelength may yield a finite temperature derivative. Crucially, as long as the measurements are taken in the linear regime, as captured by our mathematical description of the experiment (Appendix A), the phase shift is independent of the value of reflectivity and its temperature derivative. However, the noise in the phase shift is proportional to the signal-to-noise ratio of the reflectivity change $\delta \rho $; so better data quality is obtained for larger $\delta\rho $.

\subsection{\label{sec:anisotropy}In-plane inhomogeneity and anisotropy }
One of the advantages of optics-based diffusivity measurement over methods based on traditional measurements of thermal conductivity and heat capacity is the ability to perform spatially resolved measurements, enabled in our setup by mounting the sample on a 3-axis piezo-stage located in the cryostat. This allows for the measurement of diffusivity in samples which are not spatially uniform, whether due to disorder or formation of domains.

 Furthermore, this non-contact technique enables measurements of anisotropic diffusivity in a single sample, without any setup modifications or additional sample preparation. If the diffusivity along two principal axes is different ($D_\textit{\textbf{a}}$ and $D_\textit{\textbf{b}}$ for the two directions), diffusivity measured along an arbitrary direction $\theta$ is equal to: 
 \begin{equation}
 \frac{1}{D} =\frac{{\rm cos}^2\theta}{D_\textit{\textbf{a}}} +\frac{{\rm sin}^2\theta}{D_\textit{\textbf{b}}}.
 \label{Manisotropy}
\end{equation}
The derivation of the Eq. \ref{Manisotropy}, consistent with Ref. \cite{Zhang5378}, can be found in the Appendix \ref{ol}. In Ref. \cite{salazar1995novel, salazar1996thermal},  an expression for the angular dependence of the thermal diffusivity is given, but a slightly different expression is obtained due to an error in the derivation, which we explain and correct in Appendix \ref{ol}. In the experiment the angle $\theta$ corresponds to the orientation of the separation between the source and probe spots relative to \textit{\textbf{a}}-axis. By varying $\theta$ between 0 and 90 degrees, diffusivity along both principal directions can be measured, as we demonstrate below. 

\section{\label{sec:Exp}Results}
\subsection{\label{sec:level2}Studied materials}
We tested our techniques on two stoichiometric bilayer ruthenates, Ca$_{3}$Ru$_{2}$O$_{7}$ and Sr$_{3}$Ru$_{2}$O$_{7}$, as well as  Ti-doped Ca$_{3}$Ru$_{2}$O$_{7}$. Ca$_{3}$Ru$_{2}$O$_{7}$ belongs to the orthorhombic space group $Bb{\rm 2}_{1}m$, with lattice constants of $a$ = 0.5396 nm, $b$ = 0.5545 nm and $c$ = 1.961 nm \cite{yoshida2005crystal,kikugawa2010ca3ru2o7}. Single crystals exhibit twin domains and in-plane anisotropy, allowing us to use the spatial resolution to identify the domains, and measure the diffusion anisotropy, which has not been previously reported. In contrast, Sr$_{3}$Ru$_{2}$O$_{7}$  has lattice constants $a$ = $b$ = 0.3890 nm and $c$ = 2.0732 nm, and an electrical resistivity that is isotropic in the \textit{\textbf{ab}} plane, within experimental resolution, for all temperatures above 2 K \cite{shaked2000neutron,hu2010surface,mackenzie2012quantum}. We performed temperature-dependent measurements on both Ca$_{3}$Ru$_{2}$O$_{7}$ and Sr$_{3}$Ru$_{2}$O$_{7}$, to demonstrate that the two measurement configurations provide consistent results for the average in-plane diffusivity. 

The enhanced spatial resolution provided by the overlapped configuration offers opportunities to investigate the local properties, especially in doped systems which exhibit inherent disorder. To confirm this, we studied Ti-doped Ca$_{3}$Ru$_{2}$O$_{7}$. Ti doping into Ru sites in Ca$_{3}$Ru$_{2}$O$_{7}$ has been used to tune the magnetic and Mott transitions \cite{ke2011emergent,peng2013quasi}. The Ti-doped Ca$_{3}$Ru$_{2}$O$_{7}$ exhibits a metal-insulator transition (MIT), whose temperature is strongly dependent on the Ti doping level. By changing the doping level from 3$\%$ to 10$\%$, the MIT temperature varies from 50 K to 110 K. Therefore, by identifying the MIT temperature locally, we are able to reveal the local doping level, and check the in-plane chemical homogeneity of the sample.

\subsection{\label{sec:olandsep}Comparison of the overlapped and separated configuration}
To establish that the in-plane diffusivity can be measured in both configurations, it is necessary to confirm that (a) the low-frequency regime described above can be reached in both configurations, and (b) both configurations yield the same diffusivity. In FIG. 2(a) and 2(b), we show the frequency dependence of phase delay at several temperatures ranging from 300 K to 50 K in Ca$_{3}$Ru$_{2}$O$_{7}$ as a function of the square root of the modulation frequency for separated and overlapped configurations, respectively, together with the linear fits (dashed lines). The upper limit of the linear regime, marked by a grey dotted line, is 4×10$^{4}$ rad/s for separated, and 3.4×10$^{4}$ rad/s for overlapped beams. We note a spurious intercept of a few degrees at zero frequency for both configurations. The intercept is a temperature-independent setup artefact, which is easily accounted for in the analysis: if the measured frequency dependence varies as $A + B\sqrt{\omega}$, the intercept $A$ can be subtracted. This observation emphasises the importance of measuring, rather than assuming, the frequency dependence of the phase delay. The intercept subtraction is not a perfect procedure, and is likely prone to introduce errors in the cases where the absolute value of the phase delay is small, as we discuss below.  However, as long as the intercept is small compared to the frequency-dependent part of the signal, the procedure works well.

Next, we compared the diffusivity obtained from the two configurations in Ca$_{3}$Ru$_{2}$O$_{7}$ and Sr$_{3}$Ru$_{2}$O$_{7}$. In Ca$_{3}$Ru$_{2}$O$_{7}$, the diffusivity anisotropy can be measured by separating the two laser spots along two principal axes, while the overlapped measurement is sensitive to the average diffusivity (see Appendix A for more details). In FIG. \ref{Fig2}(c), we plot the diffusivity along the two principal directions measured in the separated configuration, their average value, and the result of the measurement in the overlapped configuration. The average value obtained from the separated configuration (according to Eq. \ref{Dave} in Appendix A) agrees well with the value obtained from the overlapped configuration. Furthermore, we confirm that the temperature-dependent diffusivity obtained from the two configurations in the isotropic Sr$_{3}$Ru$_{2}$O$_{7}$ is consistent (FIG. \ref{Fig3}(d)). Taken together, these results unambiguously show that the average in-plane diffusivity can be obtained by measuring the phase difference between overlapped source and probe laser beams, enabling the measurement of thermal diffusivity with diffraction-limited spatial resolution.

Although the overall agreement between the two methods is good, we note  that the discrepancy increases at low temperatures, as shown in the deviation plot in the inset of FIG. \ref{Fig3}(d). As discussed above, the error associated with the intercept subtraction procedure becomes more important as the intrinsic delay phase becomes smaller. This is most serious for the overlapped configuration at low temperatures, and likely accounts for the larger deviation from the results in the separated configuration in that temperature range. In most circumstances, the delay phase is at its smallest for any given configuration of the pump and probe pulses at low temperatures.   However, it is precisely in this low-temperature region that other methods for obtaining the diffusivity (for example a direct combination of thermal conductivity and heat capacity) are most accurate. The full power of the optical technique for our purposes comes in the region 100 - 300 K where standard thermal conductivity measurements are subjected to quite large systematic errors. Sometimes, however, even qualitative low-temperature information is useful, particularly when combined with high spatial resolution.  We will discuss such a case in the latter section of detecting diffusivity inhomogeneity.

\begin{figure*}
\includegraphics[width=6in]{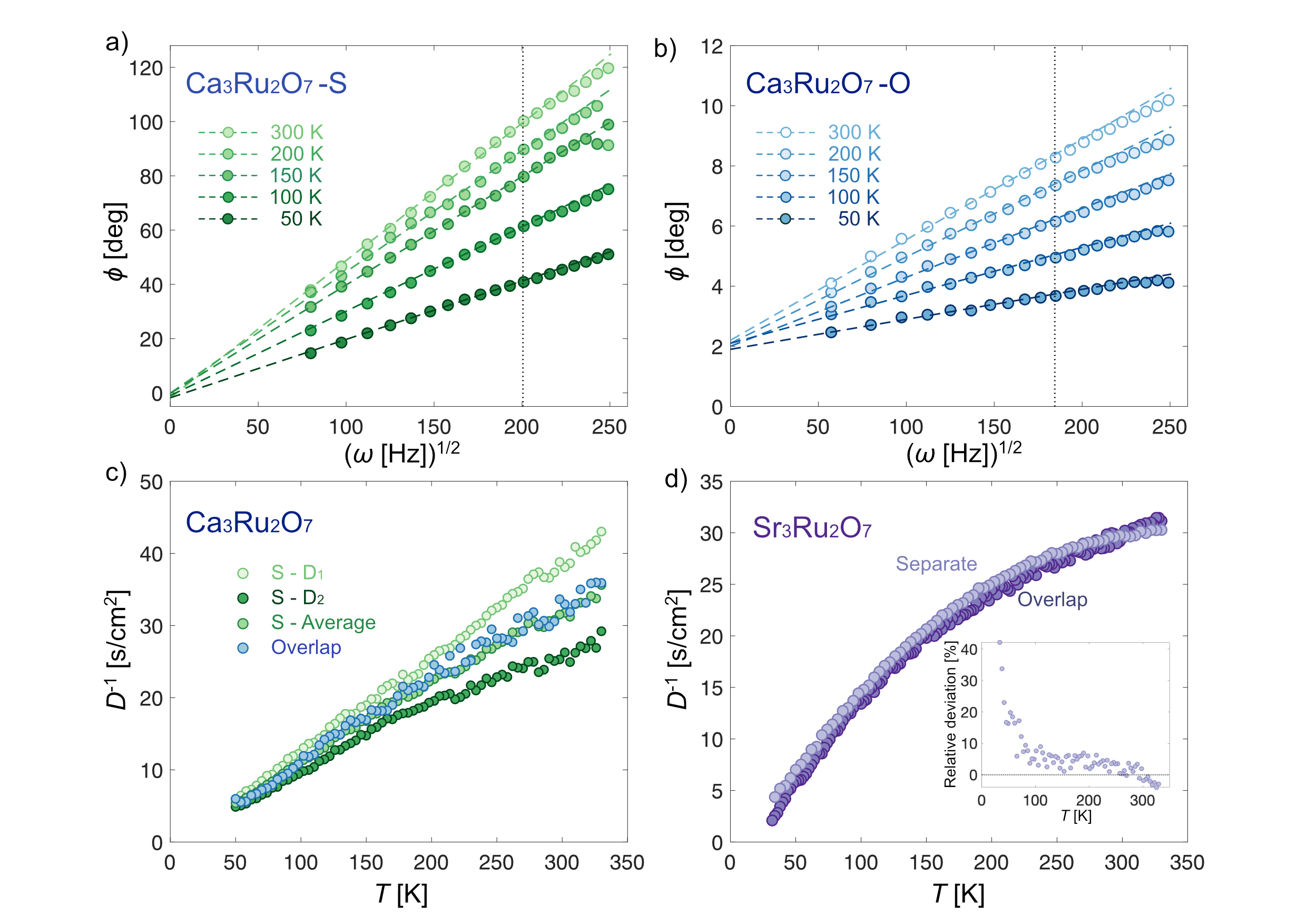}
\caption{\label{Fig2} The frequency-dependent phase delay measured in Ca$_{3}$Ru$_{2}$O$_{7}$ using the a) separated (S) and b) overlapped (O) configurations at temperatures ranging from 300 K to 50 K, plotted against the square root of the modulation frequency. The circles are the experimental data and the dashed lines are the linear fits. The vertical dotted lines estimate the upper limit of the linear regime. c) The thermal diffusivities of Ca$_{3}$Ru$_{2}$O$_{7}$, obtained from the two configurations. For separated configuration, the results in two domains ($D_1$ and $D_2$) and their average values are shown. d) The diffusivities in Sr$_{3}$Ru$_{2}$O$_{7}$, obtained from the two configurations. The inset shows the relative deviation between the results from the two configurations.}
\end{figure*}

\begin{figure*}[b]
\includegraphics[width=6in]{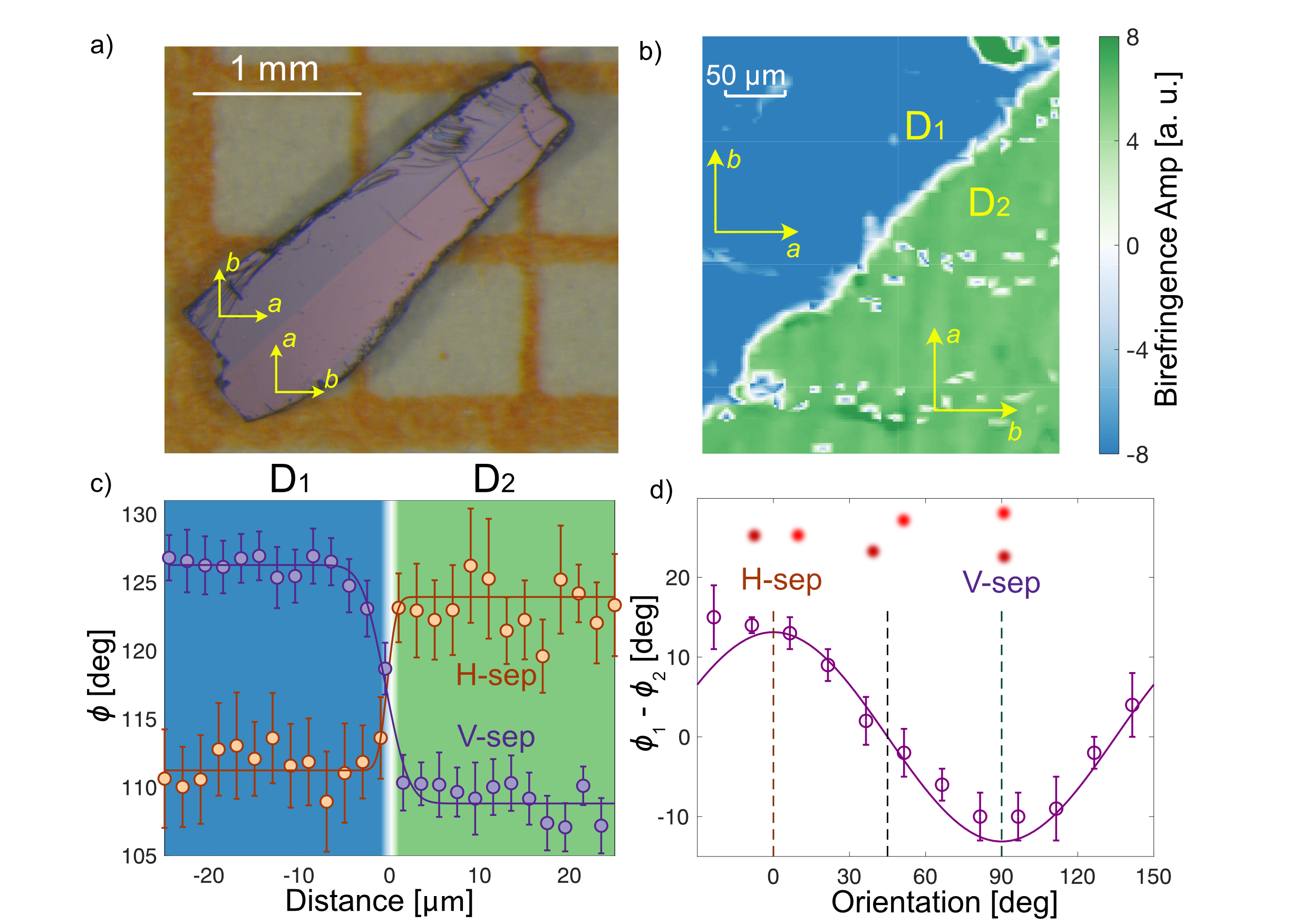}
\caption{\label{Fig3} a) Polarized optical microscopy image of Ca$_{3}$Ru$_{2}$O$_{7}$ reveals two domains. b) Optical birefringence of a part of the sample shown in a), mapped at  55K using the scanning optical setup; c) Phase difference as a function of position across the domain boundary, measured using beams separated along vertical (V-sep) and horizontal (H-sep) direction; d) Phase difference as a function of the angle between the spot separation and the horizontal axis. The solid curve is a fit to Eq.\ref{Manisotropy}. The red dots at the top of the plot are schematic illustrations of the relative orientations of the source and probe spots.}
\end{figure*}

\subsection{\label{sec:DinCRO}In-plane anisotropy of the diffusivity in Ca$_{3}$Ru$_{2}$O$_{7}$ }
By using polarized light microscopy, one can clearly see the two-level contrast of the domain structures of Ca$_{3}$Ru$_{2}$O$_{7}$ (as shown in FIG. \ref{Fig3}(a)) arising from optical birefringence \cite{CRO327SHG}. In our optical setup with a micron-scale resolution, we are able to measure spatially resolved birefringence in the same sample by employing a polarisation analysis, as shown in FIG. \ref{Fig3}(b). Since the domain boundaries are known to be along the (110) direction in Ca$_{3}$Ru$_{2}$O$_{7}$ \cite{CRO327SHG}, the observation of a diagonal domain boundary allows us to identify the principal crystalline axes as vertical and horizontal in our measurement configuration. In order to confirm that the thermal diffusivity is anisotropic, we fix the separation between two laser spots and move both of them across the domain boundary. Such one-dimensional line scans are conducted under different directions of the separation between the source and probe spots.  If the spot separation is kept fixed and vertical (horizontal), the measurement is sensitive to diffusivity $D_\textit{\textbf{b}}$ ($D_\textit{\textbf{a}}$) in domain 1 and $D_\textit{\textbf{a}}$ ($D_\textit{\textbf{b}}$) in domain 2. Performing such a measurement as a function of position along a one-dimensional line intersecting the domain boundary reveals a step at the boundary, shown in FIG. \ref{Fig3}(c) for the vertical (purple) and horizontal (orange) separation. Additionally, we can continuously rotate the spots about each other, and obtain the phase difference as a function of the angle $\theta$  between the spot separation and the principal axes (FIG. \ref{Fig3}(d)). The sinusoidal variation of the experimental results (open circles) offers additional proof that the diffusivity is indeed anisotropic. The purple curve is a fit to Eq. \ref{Manisotropy}.

\begin{figure*}[h]
\includegraphics[width=6in]{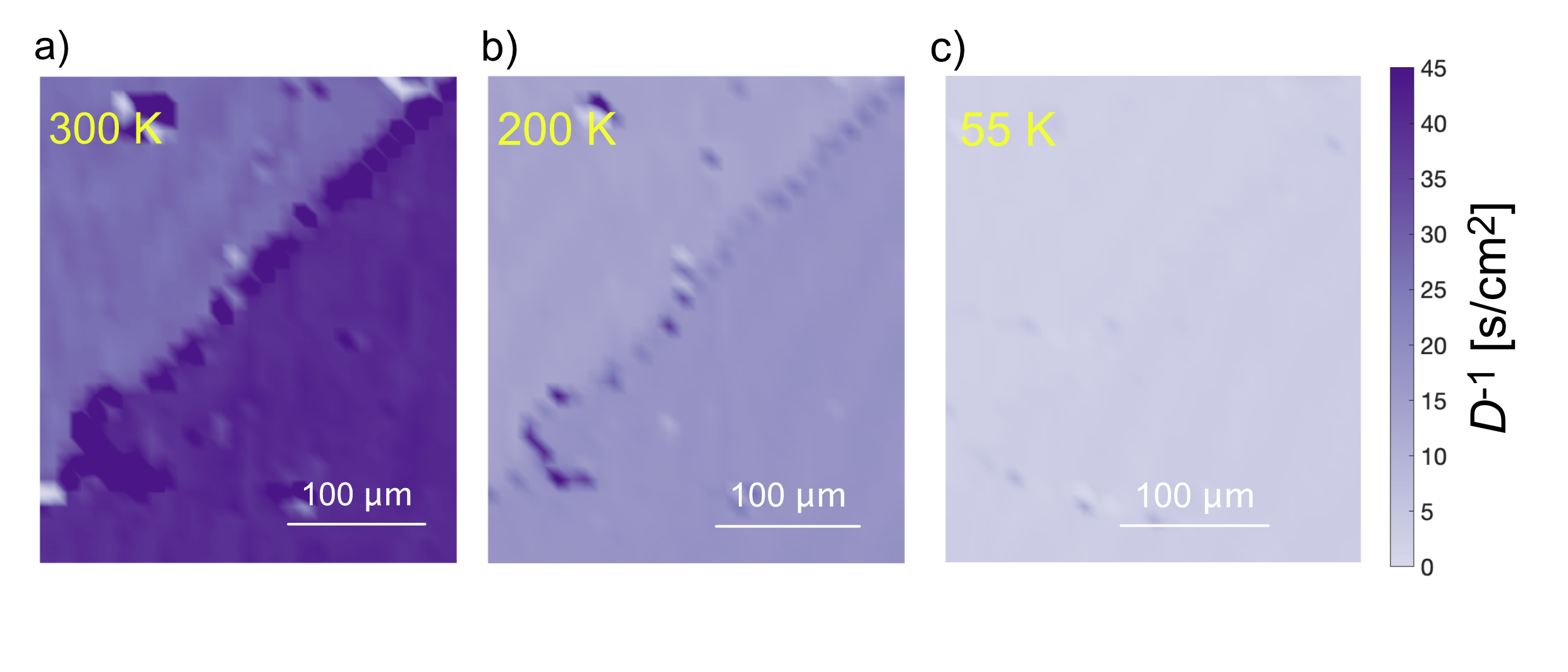}
\caption{\label{Fig4} Thermal diffusivity mapping on Ca$_{3}$Ru$_{2}$O$_{7}$ at a) \unit[300]{K}, b) \unit[200]{K} and c) \unit[55]{K}. Two domains can be clearly observed at high temperatures, indicating the anisotropy of thermal diffusivity, which decreases with decreasing temperature.}
\end{figure*}

Furthermore, our scanning setup allows us to map diffusivity over large two-dimensional regions. In FIG. \ref{Fig4}(a-c) we show the phase delay, straightforwardly related to diffusivity (Eq.~\ref{SepPhase}), across a 300 µm $\times$ 250 µm area at three selected temperatures. Two main observations can be made. Firstly, the two-domain structure is clearly identified, confirming that the thermal diffusivity is anisotropic; secondly, the anisotropy decreases with decreasing temperature. In addition, increased values of the phase delay near the domain boundary (FIG. \ref{Fig3}) are evidence of enhanced scattering at the boundary, emphasising the power of spatially resolved measurements to access the intrinsic behaviour of homogeneous domains.

\subsection{\label{sec:inhomogeneity}Detecting diffusivity inhomogeneity}
\begin{figure}[h]
\includegraphics[width=3.7in]{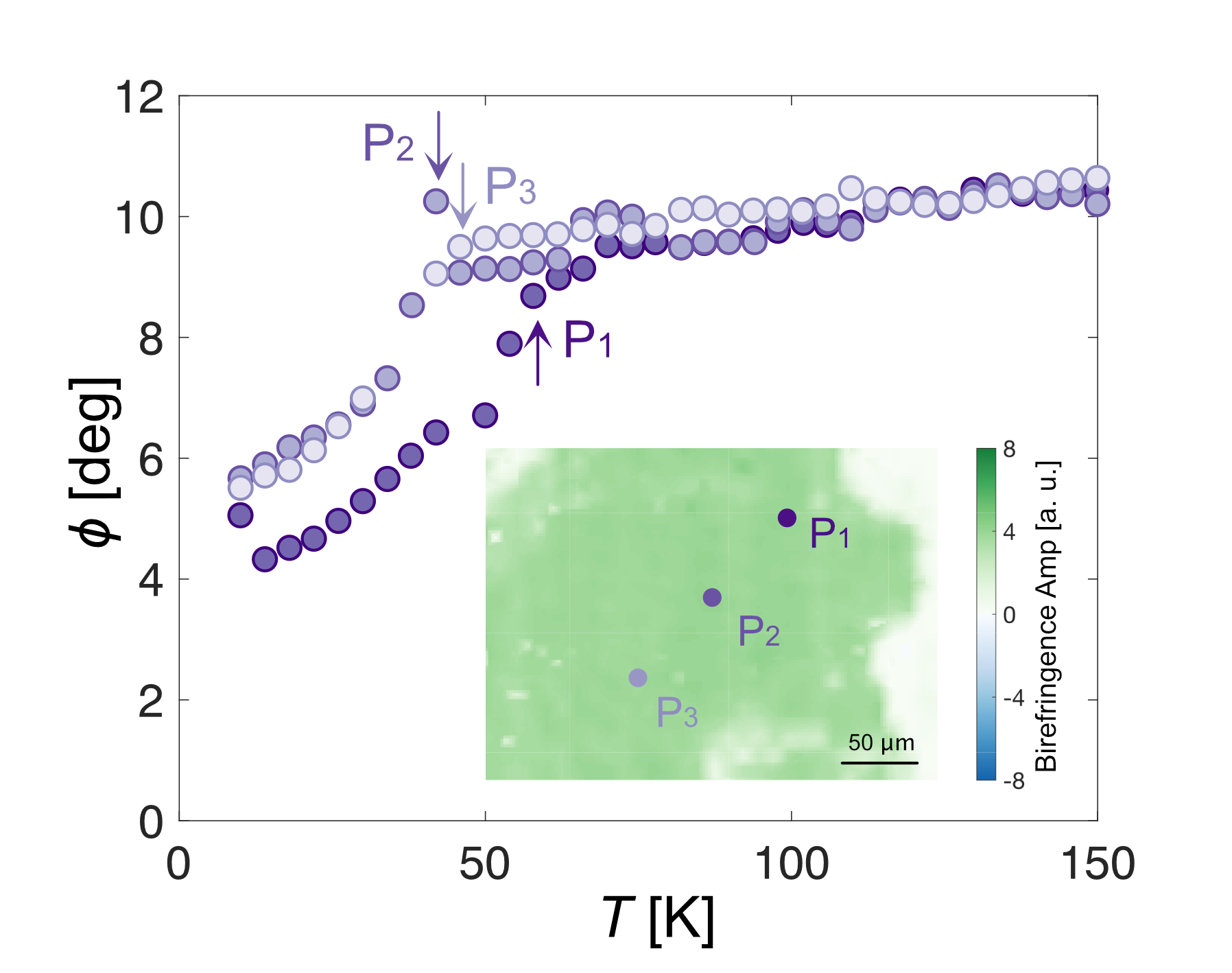}
\caption{\label{Fig5} The temperature-dependent phase difference at three different positions on the same sample of a Ti-doped Ca$_{3}$Ru$_{2}$O$_{7}$. The arrows mark the metal-insulator transition, whose temperature varies with the position on the sample. The inset is a  map of the sample birefringence, with marked measurement points P$_{1}$, P$_{2}$ and P$_{3}$, showing that the data are taken in a single domain.}
\end{figure}

The greatest advantage of the overlapped configuration is enhanced spatial resolution, limited by the beam radii which are 1.9 µm for the probe and 2.5 µm for the source in our measurements (Appendix F), but could be as small as 0.5 µm in a diffraction-limited experiment. Spatial resolution is invaluable in the study of inhomogeneous systems, such as doped samples, in which the doping level may vary across the crystal, leading to variations in physical properties, including thermal diffusivity.  In FIG. \ref{Fig5}, we show the temperature-dependent phase delay for three different positions on the same Ti-doped sample, $\rm{Ca}_{3}(\rm{Ru}_{1-\textit{x}}\rm{Ti}_{\textit{x}})_{2}\rm{O}_{7}$, with the nominal doping level of $x$ = 5$\%$ (the corresponding sample positions are marked in the birefringence inset of FIG. \ref{Fig5}). Sharp jumps in phase indicate the metal-insulator transition, which varies from \unit[50]{K} to \unit[80]{K}  for sample positions which are only $\sim\unit[70]{\mu m}$  apart. Such variations of the transition temperature across the macroscopic sample originate from the inherent doping inhomogeneity, which for this sample is comparable in magnitude to the nominal doping itself. The clear observation of the variations in transition temperature is enabled by the high spatial resolution of the overlapped configuration. It is a valuable improvement compared with traditional bulk measurements or even with the separated spot optical measurements, as both such methods unavoidably exhibit a significant broadening in the transition due to the need to average over the separation distance from pump to probe beams. In addition, the fact that the diffusivity goes up not down below the metal-insulator transition when the electrons are frozen out indicates that the diffusivity of the Ti-doped sample is dominated by phonons at high temperatures. This observation supports the aforementioned statement that even qualitative knowledge of the low-temperature diffusivity can be very useful.

\section{Conclusion}

Our results demonstrate the potential of thermal diffusivity measurement by optical methods in a cryogenic microscope setup. Separated and overlapped beam configurations provide consistent results for thermal diffusivity and exhibit their own advantages and disadvantages:  the separated configuration can be used to identify the in-plane anisotropy, while the overlapped configuration provides a higher spatial resolution, which is especially helpful when investigating potentially inhomogeneous systems. Thermal diffusivity within domains and across domain boundaries can be studied.  When working with metals, our fundamental signal, the phase shift, becomes smaller as the temperature decreases and the diffusivity increases.  This means that the accuracy of the optical method decreases at low temperatures.  However, this fact provides a useful complementarity with traditional methods of deducing the diffusivity from measurements of thermal conductivity and heat capacity, because the accuracy of traditional thermal conductivity measurements is poorest at high temperatures.  We believe the combination of the two methods will pave the way to measurement over several decades of temperature, yielding higher quality information than is currently available.  More broadly, our setup gives us the possibility of studying probes of symmetry breaking, such as birefringence, Kerr rotation, second harmonic generation, \textit{etc}. with the same spatial resolution as we have demonstrated here for thermal diffusivity. The combination will allow us to explore and visualize thermal properties in broken symmetry phases of correlated materials.

\begin{acknowledgments}
We wish to acknowledge Aharon Kapitulnik and Jiecheng Zhang for useful discussions, and for bringing independent work of Ref. \cite{zhang2020optical} to our attention. We also acknowledge Elena Hassinger and Ulrike Stockert for their useful discussions. VS is
supported by the Miller Institute for Basic Research
in Science, University of California, Berkeley. NK is supported by a KAKENHI Grants-in-Aids for Scientific Research (Grant Nos. 17H06136, 18K04715, and 21H01033), and Core-to-Core Program (No. JPJSCCA20170002) from the Japan Society for the Promotion of Science (JSPS) and by a JST-Mirai Program (Grant No. JPMJMI18A3).  APM and SM acknowledge the financial support of the Deutsche Forschungsgemeinschaft (DFG, German Research Foundation) - TRR 288  -422213477 (project A10). Research in Dresden benefits from the environment provided by the DFG Cluster of Excellence ct.qmat (EXC 2147, project ID 390858940).
\end{acknowledgments}

\section*{Author declaration}
\subsection*{Conflicts of interests}
The authors have no conflicts to disclose.

\subsection*{Author Contributions}
\textbf{FS} Conceptualization(equal); Formal analysis (equal); Investigation(equal); Writing –
original draft (equal); Writing - review \& editing(equal);
\textbf{SM} Conceptualization(supporting); Formal analysis (equal); Investigation(equal); Writing –
original draft (supporting); Writing - review \& editing(supporting);
\textbf{PHM} Investigation(supporting); Writing - review \& editing(supporting);
\textbf{ZHF} Investigation(supporting); Writing - review \& editing(supporting);
\textbf{IM} Investigation(supporting); Writing - review \& editing(supporting);
\textbf{DAS} Investigation(supporting); Writing - review \& editing(supporting);
\textbf{NK} Investigation(supporting); Writing - review \& editing(supporting);
\textbf{JWO} Conceptualization(supporting); Writing - review \& editing(supporting);
\textbf{SAH} Conceptualization(equal); Investigation(supporting); Writing - review \& editing(supporting);
\textbf{APM} Conceptualization(equal); Supervision (equal); Writing –
original draft (equal); Writing - review \& editing(equal);
\textbf{VS} Conceptualization(equal); Investigation(equal); Supervision (equal); Writing –
original draft (equal); Writing - review \& editing(equal);

\section*{Data Availability Statement}
The data that support the findings of this study are available from the corresponding author upon reasonable request.


\clearpage
\appendix
\setcounter{equation}{0}
\renewcommand\theequation{A.\arabic{equation}}
\setcounter{figure}{0}
\renewcommand\thefigure{A.\arabic{figure}}
\section{Solution of the diffusion equation}
The diffusive transport is governed by the diffusion equation
\begin{equation}
 \frac{\partial{\Phi}}{\partial t}-
 (D_x\frac{\partial^2{\Phi}}{\partial x^2}+D_y\frac{\partial^2{\Phi}}{\partial y^2}+D_z\frac{\partial^2{\Phi}}{\partial z^2}) = \frac{{I_s}}{c},
 \label{Adiffequ}
\end{equation}
$\Phi (x,y,z,t)$ is the temperature profile; $I_s(x,y,z,t)$ is the external heat source; $c$ is the heat capacity; $D_x$, $D_y$ and $D_z$ are the thermal diffusivities along three principal directions, when they are coincident with the principal axes of the thermal diffusivity tensor. The same equation describes both of our experimental configurations; the key difference is that the source term $I(x,y,z,t)$ can be approximated as a delta function  in the separated configuration, but has to be described as a Gaussian in the overlapped case.

We assume that with a small temperature fluctuation $\delta T$, both heat capacity and thermal diffusivity stay constant. We use Green’s function method to solve Eq.\ref{Adiffequ}. Firstly, we Fourier transform the temperature distribution and the source term:
\begin{equation}
 \Phi(x,y,z,t)= \int \Phi_k (k_x,k_y,k_z,t)e^{-i\boldsymbol{kr}}d\boldsymbol{k},
 \label{AFTphi}
 \end{equation}
 
 \begin{equation}
 I_s(x,y,z,t)= \int I_{sk}(k_x,k_y,k_z,t)e^{-i\boldsymbol{kr}}d\boldsymbol{k},
 \label{AFTI}
 \end{equation}
Using the Fourier transform $\Phi_k$ and $I_{sk}$, Eq.\ref{Adiffequ} becomes:

\begin{equation}
 (\frac{\partial}{\partial t}+{D_x}{k_x^2}+{D_y}{k_y^2}+{D_z}{k_z^2})\Phi_k =  \frac{{I_{sk}}}{c},
 \label{AdiffequinK}
\end{equation}
where $k_x$, $k_y$ and $k_z$ are the three basis vectors in momentum space. The Green’s function of the operator $(\frac{\partial}{\partial t}+{D_x}{k_x^2}+{D_y}{k_y^2}+{D_z}{k_z^2})$ in the momentum space can be expressed as:
\begin{equation}
 G(k_x,k_y,k_z,t)= \Theta (t)e^{-({D_x}{k_x^2}+{D_y}{k_y^2}+{D_z}{k_z^2})t},
 \label{Greenk}
 \end{equation}
The Green’s function in real space is obtained by the inverse Fourier transform. Taking into account the boundary condition at the surface, which prohibits heat flow out of the material, we obtain:
\begin{widetext}
    \begin{equation}
    G(x,y,z,t,z')=-\frac{\Theta (t)}{{2\pi}^3} \int dk_x dk_y dk_z e^{i({k_x}x+{k_y}y)} {\rm cos}(k_zz){\rm cos}(k_zz{'}) e^{-({D_x}{k_x^2}+{D_y}{k_y^2}+{D_z}{k_z^2})t},
     \label{Greenr2}
     \end{equation}
Note the explicit dependence on $z$ and $z'$, caused by the breaking of translational symmetry at the sample surface.With the Green's function, we can obtain the solution to Eq. \ref{Adiffequ}.
 \begin{equation}
 \Phi(x,y,z,t)= \int dt{'}dx{'}dy{'}dz{'} G(x-x',y-y',t,z,z{'}) \frac{{I_k(x{'},y{'},z{'},t{'})}}{c},
 \label{PhiSolu}
 \end{equation}
The transient change in reflectivity $\delta R$ we measure is proportional to the local transient change in temperature $\delta T$, obtained by integrating the temperature distribution $\phi(x,y,z,t)$ over the probed region:
\begin{equation}
 \delta T \propto \int dtdxdydz\Phi(x,y,z,t)I_p(x,y,z,t),
 \label{Atemp}
\end{equation}
All of the expressions derived above are valid for both measurement configurations; the difference arises due to different relative position of the source and probe beams, as we describe below.

\subsection{\label{ol}Overlapped configuration}
Both the probe and the source take the Gaussian form:
\begin{equation}
 I_s (x,y,z,t) \propto e^{-(x^2+y^2)/{2R_s^2}}e^{-z/z_s}e^{-i\omega t},
 \label{Asource}
\end{equation}
\begin{equation}
 I_p (x,y,z,t) \propto e^{-(x^2+y^2)/{2R_p^2}}e^{-z/z_p}e^{-i\omega ' t},
 \label{Aprobe}
\end{equation}
where $R_s$ and $R_p$ are the radius of the source and probe beam, respectively; $z_s$ and $z_p$ are the penetration depth: we assume $z_s = z_p$.
we can now perform the integral over the positions in Eq. \ref{Atemp} and obtain:
\begin{equation}
 \delta T \propto \int d^3k \frac{e^{-(k_x^2+k_y^2)(R_s^2+R_p^2)}} {(1+k_z^2 z_s^2)^2} \frac{1}{i\omega -({D_x}{k_x^2}+{D_y}{k_y^2}+{D_z}{k_z^2})},
 \label{Atemp2}
\end{equation}
This expression must in general be evaluated numerically. In the special case of isotropic two-dimensional diffusion ($D_x=D_y=D, z_s \rightarrow 0$) it can also be solved analytically:
\begin{equation}
 \delta T \propto \int d^3k e^{\frac{i(R_s^2+R_p^2)\omega}{2D}} {\rm Erfc}(\sqrt{-\frac{i(R_s^2+R_p^2)\omega}{2D}})
 \label{Atemp3}
\end{equation}
In the low-frequency limit ($\omega \ll \frac{D}{(R_s^2+R_p^2)}$)
, we find the phase offset between the heat and the temperature change ($\phi = {\rm Arg}(\delta T)$) to be:
\begin{equation}
 \phi = \sqrt{\frac{(R_s^2+R_p^2)\omega}{\pi D}}
 \label{Olphi}
\end{equation}

Worthy to note that, for the anisotropic case where $D_x \neq D_y$, the diffusivity $D$ we obtained from Eq. \ref{Olphi} is the average diffusivity $D_{\rm ave}$ of the two in-plane directions $D_x$ and $D_y$. In this case, the phase offset can also be extracted from Eq. \ref{Atemp2} and given by an elliptic integral:
\begin{equation}
 \phi = \sqrt{\frac{\pi(R_s^2+R_p^2)\omega}{4}} \left[ \int_{0}^{\frac{\pi}{2}} \left(\frac{{\rm cos}^2\theta}{D_y}+\frac{{\rm sin}^2\theta}{D_x} \right) ^{-\frac{1}{2}}d\theta \right] ^{-1}
 \label{Olphi2}
\end{equation}
Therefore, comparing Eq. \ref{Olphi} with Eq. \ref{Olphi2}, we obtain the average diffusivity, which is the value we measure from the overlapped configuration for an in-plane anisotropic system:
\begin{equation}
 D_{\rm ave}^{\frac{1}{2}} = \frac{2}{\pi}\int_{0}^{\frac{\pi}{2}}\left(\frac{{\rm cos}^2\theta}{D_y}+\frac{{\rm sin}^2\theta}{D_x} \right) ^{-\frac{1}{2}}d\theta
 \label{Dave}
\end{equation}
We demonstrate experimentally that the approximation of 2-dimensional diffusivity is valid in Ca$_{3}$Ru$_{2}$O$_{7}$.

\subsection{\label{ol}Separated configuration}
It is clear that the experiment performed in the separated configuration probes in-plane diffusion, and the extent of the beams perpendicular to the plane can be neglected ($z_s=z_p=0$). Both beams can be approximated as delta functions:
\begin{equation}
 I_s (x,y,z,t) \propto \delta (x-x_s)\delta (y-y_s)\delta (z)e^{-i\omega t},
 \label{Asource2}
\end{equation}

\begin{equation}
 I_p (x,y,z,t) \propto \delta (x-x_p)\delta (y-y_p)\delta (z)e^{-i\omega{'} t},
 \label{Aprobe2}
\end{equation}
The change of temperature is now:
\begin{equation}
 \delta T \propto \int d^3k e^{i(k_x(x-x')+k_y(y-y'))} \frac{\delta (\omega + \omega{'})}{i\omega -{D_x}{k_x^2}+{D_y}{k_y^2}+{D_z}{k_z^2}},
 \label{Atemp5}
\end{equation}
The integral can be performed analytically:
\begin{equation}
 \delta T \propto \int \frac{kdk}{\sqrt{k^2-i\omega}}J_0 \left(k\sqrt{\frac{\Delta x^2}{D_x}+\frac{\Delta y^2}{D_y}}\right)
 \label{Atemp6}
\end{equation}
\end{widetext}
where $\Delta x$ and $\Delta y$  represent the separation between two spots along two directions, and $J_0$ is the zero order Bessel function of the first kind. From Eq. \ref{Atemp6}, we obtain the phase difference between the source and probe beams:
\begin{equation}
 \phi = \sqrt{\frac{\omega}{2}\left(\frac{\Delta x^2}{D_x}+\frac{\Delta y^2}{D_y}\right)}
 \label{SepPhi}
\end{equation}
If the system is in-plane isotropic, we have $D_x$ = $D_y$ = $D$, them the Eq. \ref{SepPhi} becomes:
\begin{equation}
 \phi = \sqrt{\frac{r^2\omega}{2D}}
 \label{SepPhi2}
\end{equation}
where $r^2= {\Delta x}^2+{\Delta y}^2$. Eq. \ref{SepPhi2} is used to measure the diffusivity in the separated configuration (Eq. \ref{SepPhase}  in the main text). In addition, if we compare Eq. \ref{SepPhi} and \ref{SepPhi2}, we can obtain the relation between the average diffusivity $D$ and the uniaxial diffusivities ($D_x$ and $D_y$), which is Eq. \ref{Manisotropy} in the main text.

The angular dependence of the diffusivity (Eq. \ref{Manisotropy}) is consistent to the expression in Ref. \cite{Zhang5378}. However, we noticed that there is a disagreement between our expression and that in Ref. \cite{salazar1995novel, salazar1996thermal}, which arises due to a mistake in the derivation used in those references, as we demonstrate below. Their starting point is the thermal conductivity tensor in its quadratic representation (Equation 3 in Ref. \cite{salazar1996thermal}):
\begin{equation}
 \kappa_xx^2+\kappa_yy^2+\kappa_zz^2 = 1
 \label{quadratic}
\end{equation}
This equation, defined as ‘thermal conductivity ellipsoid’,  is consistent with Ref. \cite{nye1985physical}, also cited in Ref. \cite{salazar1996thermal} as  Ref. 12. However, because the semiaxes of the ellipsoid are $\sqrt{\frac{1}{\kappa_x}}$, $\sqrt{\frac{1}{\kappa_y}}$ and $\sqrt{\frac{1}{\kappa_z}}$, respectively, for an
arbitrary direction ($\alpha$, $\beta$, $\gamma$), one should have:
\begin{equation}
 \left(\sqrt{\frac{1}{\kappa_{\alpha\beta\gamma}}}\right)^2 = \left(\sqrt{\frac{1}{\kappa_{x}}}\rm cos\alpha \right)^2+\left(\sqrt{\frac{1}{\kappa_{y}}}\rm cos\beta \right)^2+\left(\sqrt{\frac{1}{\kappa_{z}}}\rm cos\gamma \right)^2,
 \label{quadratic2}
\end{equation}
which is in contradiction with Eq. 4 in Ref. \cite{salazar1996thermal}. 

Since heat capacity is not a directional quantity, thermal diffusivity obeys the same angular dependence as thermal conductivity, leading to expression Eq. \ref{Manisotropy} of the main text for diffusivity anisotropy in 2D.

\section{Penetration depth}
\begin{figure*}
\includegraphics[width=6in]{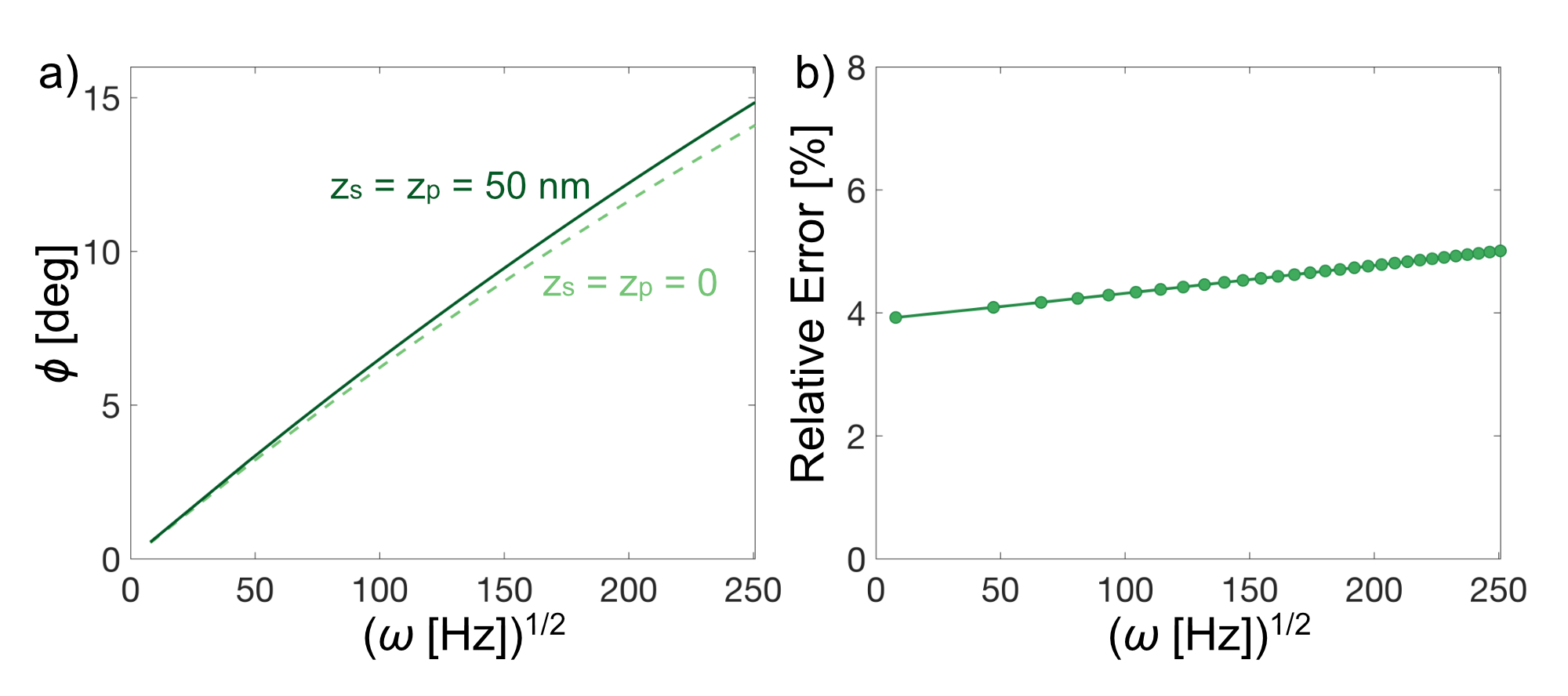}
\caption{\label{FigA1} a) The numerical solutions of the phase delay based on Eq. \ref{Atemp2}. The solid curve is calculated for a penetration depth of 50 nm for both source and probe beam, and the dashed curve for zero penetration depth. b) The relative error between two cases in a).}
\end{figure*}

As discussed in the main text, it is not immediately obvious whether an experiment performed under the overlapped configuration dominantly probes the in-plane or out-of-plane diffusion. Here we give numerical solutions to Eq. \ref{Atemp2}, based on realistic parameters: $R_s= R_p=2.3$ µm, $D= D_x= D_y$ = 5×10$^{-6}$ ${\rm m}^2/{\rm s}$, and $D_z=0.01D$. In FIG. \ref{FigA1}(a), we show the phase delay as a function of the square root of the frequency. The solid green curve is the numerical solution with a finite penetration depth $z_s= z_p=50$ nm, and the dashed curve is the solution with zero penetration depth. Small deviation can be seen lower especially at lower frequencies. In FIG. \ref{FigA1}(b), we plot the relative error between the numerical solution and the 2D approximation, and find that the maximum error is  4 - 5 $\%$. The numerical solutions indicate that even under the overlapped configuration, the phase delay between the source and the probe is dominated by the in-plane diffusion.

\section{Calibration of the focus}
\begin{figure*}
\includegraphics[width=6in]{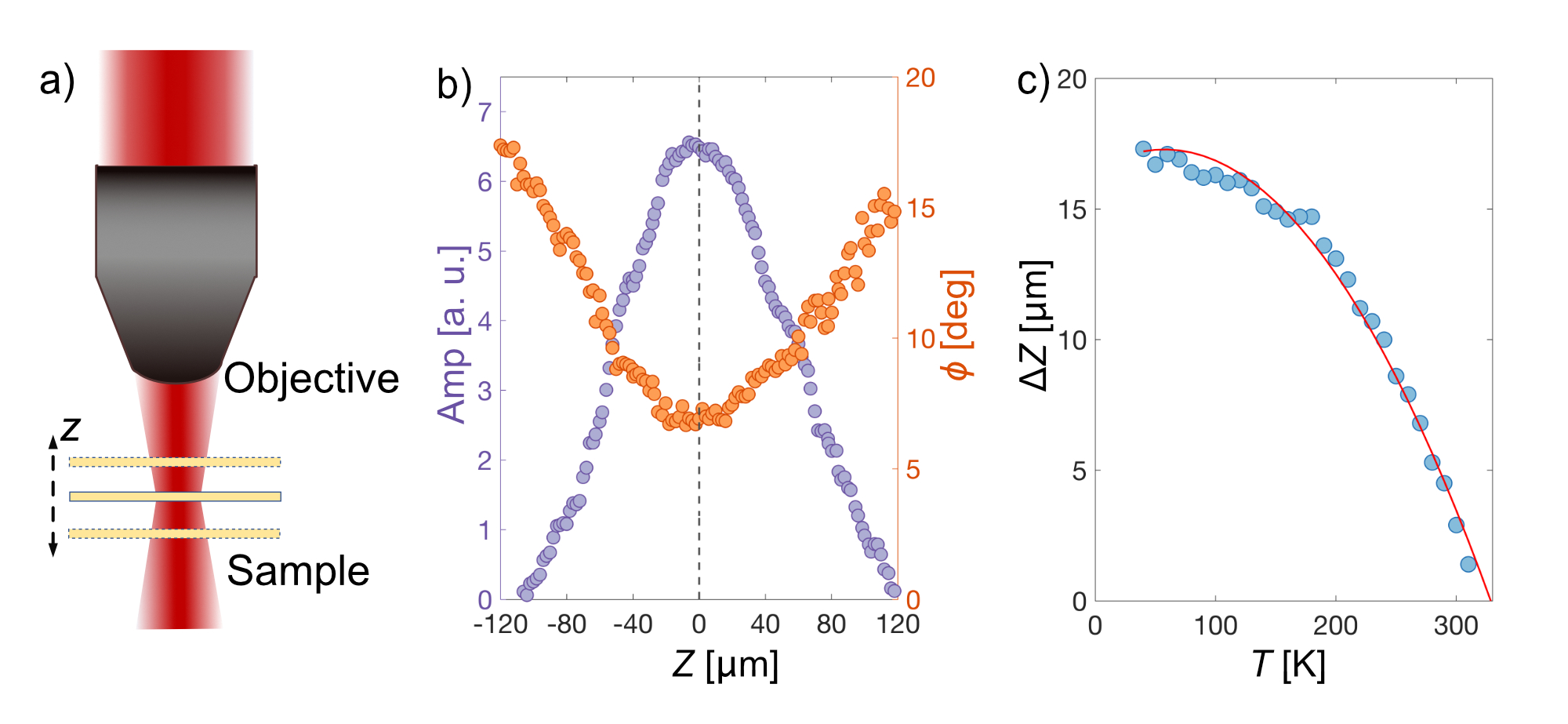}
\caption{\label{FigA2} Calibration of the $z$-shift during temperature change. a) The schematic of the $z$-scan, illustrating the change of sample position in the focused beam profile. b) The amplitude (purple) and phase (orange) of the signal as a function of sample height. The gray dashed line marks the position where the sample is in the beam focus.  c) The calibration of the stage shift along $z$ direction during a temperature ramp. The red curve is a polynomial fit.}
\end{figure*}

When performing the temperature-dependent measurements, special attention should be paid to the shift of the sample height due to the thermal expansion. In the overlapped configuration, a small $z$-shift changes the size of the laser spots, leading to a noticeable change in the phase delay (as shown in FIG. \ref{FigA2}(a)). To compensate for such shift, we firstly scan the stage along $z$-direction and record the temperature-modulated amplitude of the reflectivity, as well as the phase delay. As shown in FIG. \ref{FigA2}(b), at the focal position the amplitude is maximised and the phase minimised. By performing such $z$-scans, we are able to determine the focus with the accuracy of 1 µm. Furthermore, by measuring the $z$-scan as a function of temperature, we are able to quantify the change of the focal position due to the thermal expansion of the stage (FIG. \ref{FigA2}(c)). Once measured, such a shift can be compensated by using the piezo stage, allowing us to obtain a reliable measurement of the phase delay.

\section{Two ways of controlling the position of the laser spots}
\begin{figure*}
\includegraphics[width=6in]{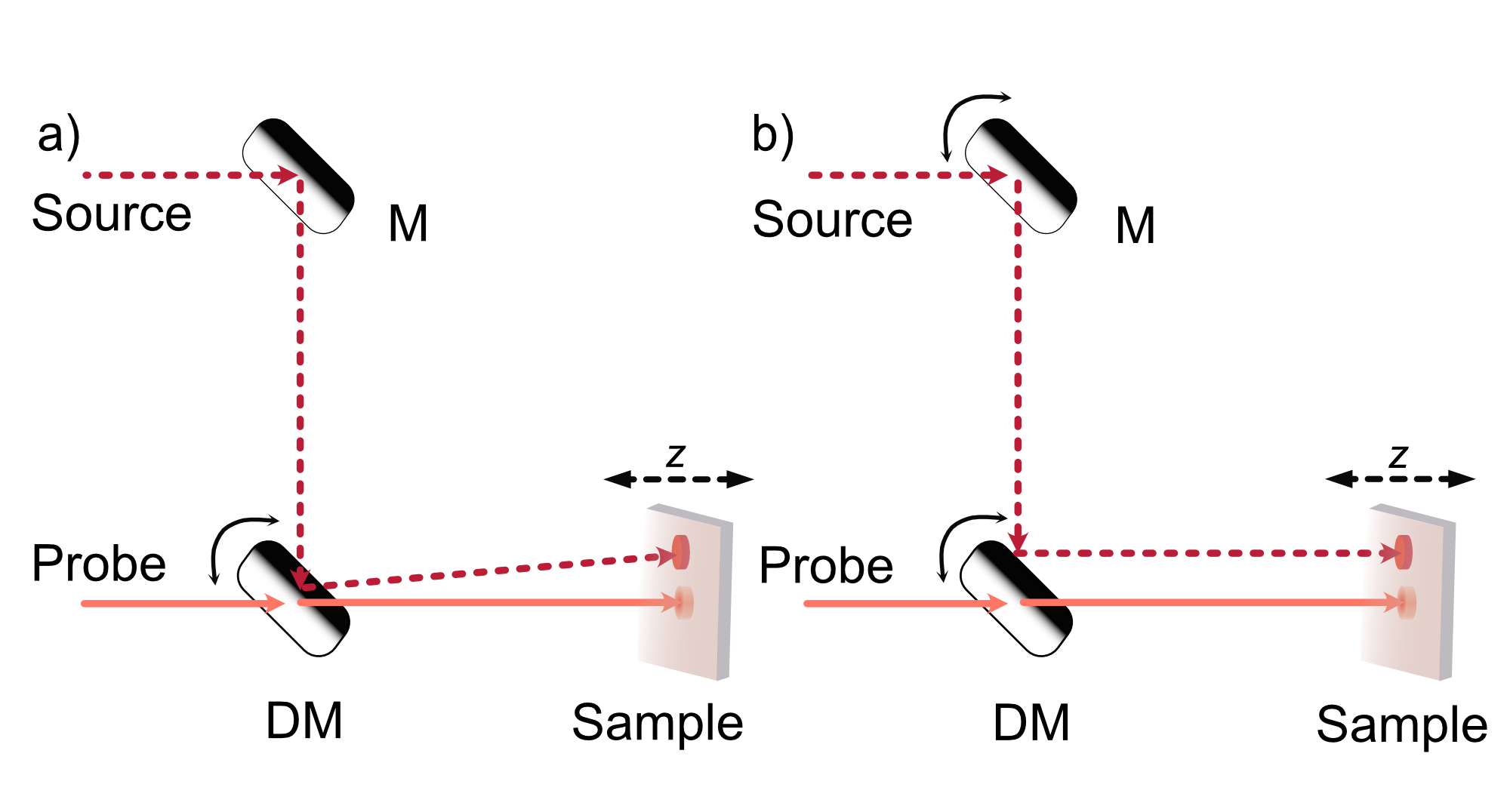}
\caption{\label{FigA3} Two methods of separating the laser beams. a) Moving source beam by adjusting dichroic mirror (DM) and keeping probe beam fixed. It is easier to operate but the separation between two beams is sensitive to sample position shift; b) Moving the source beam by coordinating M and DM together, keeping both beams normal to the sample surface. The separation is more stable against a sample position shift.}
\end{figure*}

In the separated configuration, a small change in the laser spot size does not affect the phase because the diameter of both source and probe laser is much smaller than the separation between the two spots. However, if the laser paths are not parallel to each other, the separation between the two spots can vary with the shift in the $z$-direction. In practice, for the convenience of the detection, we keep the probe beam path fixed, and normal to the sample surface. The source beam can be moved in two ways (FIG. \ref{FigA3}). The first method (FIG. \ref{FigA3}(a)) is by adjusting the dichroic mirror (DM), which allows us to continuously move the source position on the sample. It is useful and convenient when we want to change the separation or switch the configuration between overlapped and separated cases at a fixed temperature. In addition, we can also use a piezo-electrical mirror mount to control the dichroic mirror, realizing an automatic scan of the separation. However, the disadvantage is that since the source path is tilted, the separation between spots changes during a temperature-dependent measurement due to the shift of the sample along the $z$-direction. Thus, we utilized a second way to control the beam separation (FIG. \ref{FigA3}(b)), by coordinating two mirrors (M and DM), in order to control both the beam position and the incident angle. The separation is now more stable with respect to small changes in $z$.

\section{The linearity regime of $\delta R$}

\begin{figure*}[h]
\includegraphics[width=6in]{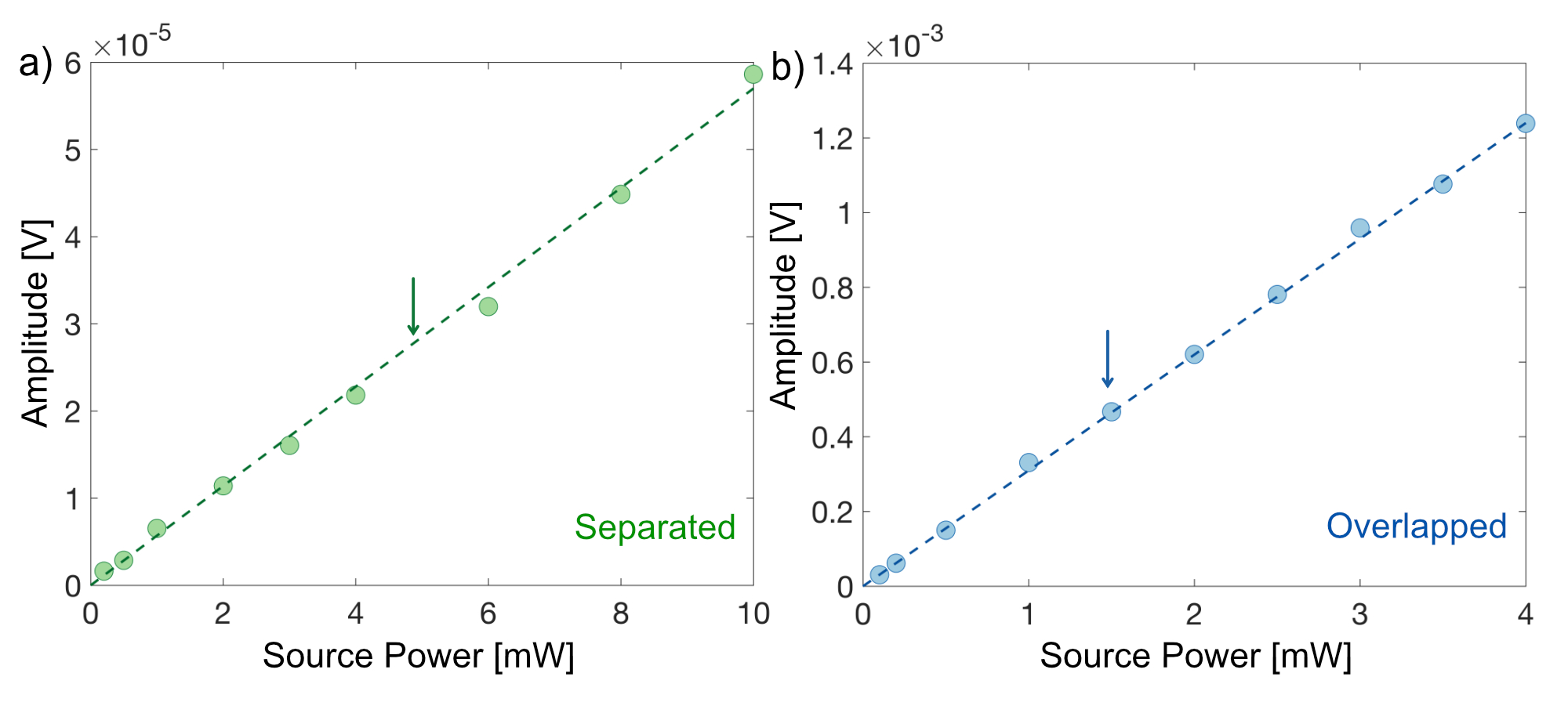}
\caption{\label{FigA4} Source power-dependent signal in Ca$_{3}$Ru$_{2}$O$_{7}$ for (a) separated and (b) overlapped  configuration. The amplitude (in unit of V) is directly measured by lock-in amplifier at the modulation frequency, and represents the differential reflectivity. The source powers we used in the diffusivity experiment are marked by arrows respectively, which are both within the linear range.}
\end{figure*}

As discussed in the main text, source laser power has to be carefully chosen.The temperature change $\delta T$ induced by it should be large enough to yield a measurable $\delta R$ signal, but small enough to ensure that  the measurement is taken in the linear regime, where  $\delta R \sim \delta T$. In order to establish the latter, the dependence of $\delta R$ on the source power should be measured under the experimental conditions (material, configuration, frequency, temperature) before each diffusivity measurement. Results of such measurements for both separated and overlapped configurations in Ca$_{3}$Ru$_{2}$O$_{7}$ are shown in FIG.\ref{FigA4}, where perfect linear dependence of $\delta R$ on source power is observed for both configurations. The source powers we used in the experiments are marked by arrows, and are both within the linear regime.

\section{Determination of spots size and the separation}

\begin{figure*}[h]
\includegraphics[width=7in]{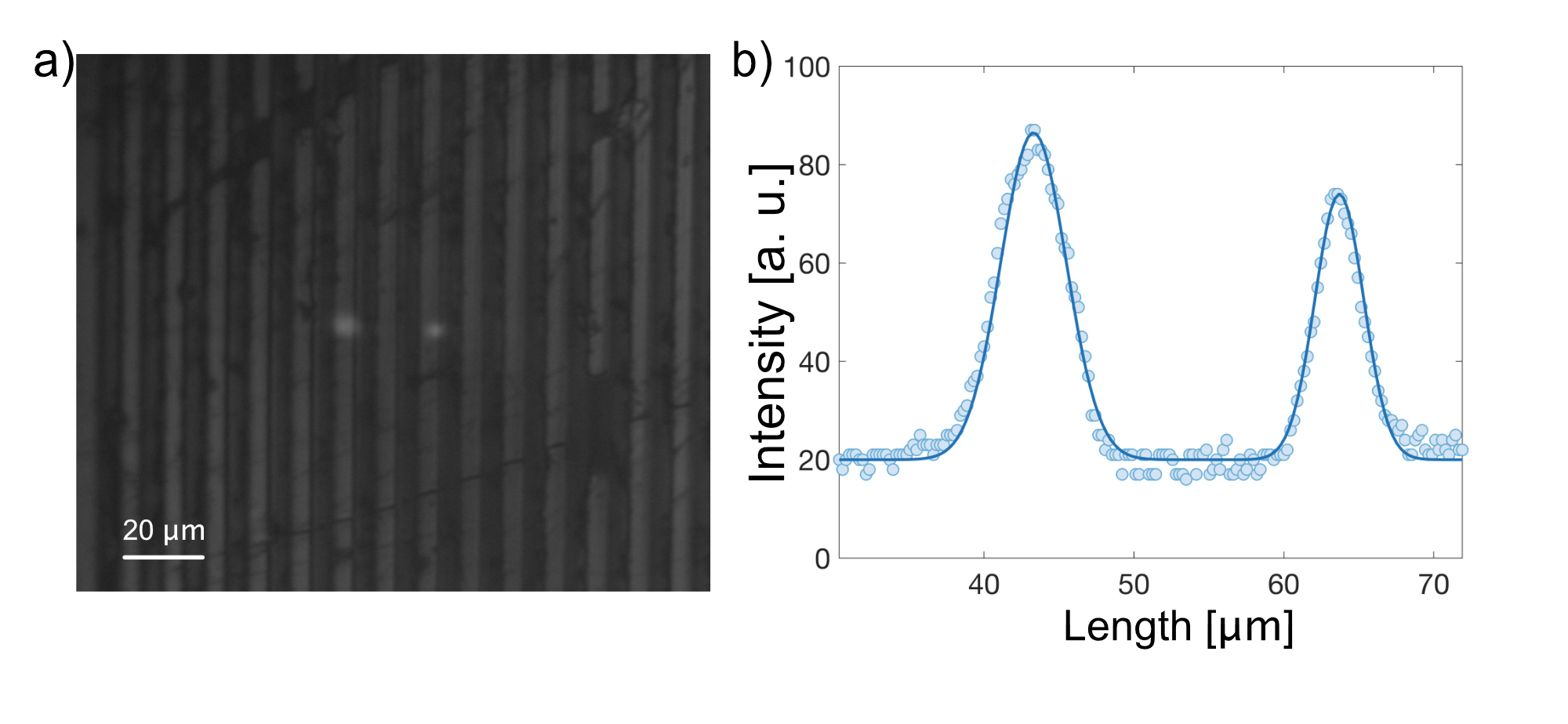}
\caption{\label{FigA5} Determination of spots size and the separation. (a) is the CCD image of the FIB calibration sample and the laser spots. The left one is the source spot and the right one is the probe spot; (b) is one example of the Gaussian fit to the laser beam profiles. The solid curve is the fitting curve, which yields radii of 2.5 µm for source, 1.9 µm for probe, and a separation of 20.6 µm.}
\end{figure*}

In order to precisely determine the spot sizes and the separation, a micro-structured calibration sample has been made by focused ion beam (FIB), which has a series of gold coated (bright) and non-gold coated (dark) stripes with the interval of 5 µm, as shown in FIG.\ref{FigA5} (a). By taking a CCD image of the calibration stripes and the laser spots simultaneously, the distance can be directly measured. To be more precise, a Gaussian function fitting can be done to analyse the profile of the laser beams. One example has been shown in FIG.\ref{FigA5} (b), where the radii of the source and probe beams are determined to be 2.5 µm and 1.9 µm respectively and the separation is 20.6 µm.

\clearpage
\nocite{*}
\bibliography{aipsamp}

\end{document}